\documentclass[10pt,twocolumn,aps,prl,superscriptaddress,reprint]{revtex4-1}

\usepackage[latin9]{inputenc}
\usepackage{bm}
\usepackage{amsmath}
\usepackage{graphicx}
\usepackage[unicode=true,
 bookmarks=false,
 breaklinks=false,pdfborder={0 0 1},colorlinks=false]
 {hyperref}
\usepackage{breakurl}

\usepackage{float}
\usepackage{threeparttable} 
\usepackage{multirow}
\usepackage{ulem}

\makeatletter

\usepackage{xcolor}
\usepackage{bm}
\usepackage{url}

\makeatother

\newcommand{\be}{\begin{equation}}
\newcommand{\ee}{\end{equation}}

\begin{document}

\title{Electron-phonon coupling and the Coexistence of Superconductivity and Charge-Density Wave  in Monolayer NbSe$_2$}

\author{Feipeng Zheng}
\email{fpzheng_phy@163.com}
\affiliation{Siyuan Laboratory, Guangzhou Key Laboratory of Vacuum Coating Technologies and New Energy Materials, Department of Physics, Jinan University, Guangzhou 510632, China}

\author{Ji Feng}
\email{jfeng11@pku.edu.cn}
\affiliation{International Center for Quantum Materials, School of Physics, Peking
University, Beijing 100871, China}
\affiliation{Collaborative Innovation Center of Quantum Matter, Beijing 100871,
China}
\affiliation{CAS Center for Excellence in Topological Quantum Computation, University of Chinese Academy of Sciences, Beijing 100190, China}

\date{\today}
\begin{abstract}
Monolayer 2$H$-NbSe$_2$ has recently been shown to be a 2-dimensional superconductor, with a coexisting charge-density wave (CDW). As both phenomena are intimately related to electron-lattice interaction, a natural question is how superconductivity and CDW are interrelated through electron-phonon coupling (EPC), which is important to the understanding of 2-dimensional superconductivity. This work investigates the superconductivity of monolayer NbSe$_2$ in CDW phase using the anisotropic Migdal-Eliashberg formalism based on first principles calculations. The mechanism of the competition between and coexistence of the  superconductivity and CDW is  studied in detail by analyzing  EPC. It is found that the intra-pocket scattering is related to superconductivity, leading to almost constant value of superconducting gaps on parts of the Fermi surface. The inter-pocket scattering is found to be responsible for CDW, leading to  partial or full bandgap on the remaining Fermi surface.  Recent experiment indicates that there is transitioning from regular superconductivity in thin-film NbSe$_{2}$ to two-gap superconductivity in the bulk, which is shown here to have its origin in the extent of Fermi surface gapping of $\mathbf{K}$ and $\mathbf{K'}$ pockets induced by CDW. Overall blue shifts of the phonons and sharp decrease of  Eliashberg spectrum are found when the CDW forms.
\end{abstract}
\maketitle
Recently, monolayer niobium diselenide in 2$H$ phase (NbSe$_2$ henceforth)  has drawn experimental and theoretical interests, as a  2-dimensional superconductor with coexisting CDW\citep{Xi2015,ugeda2016,Wang2017a,Xing2017,cao2015quality,silva2016electronic,zheng2017charge,Lian2018}. However, the mechanism of the competition and coexistence between the superconductivity and CDW remains to be clarified, both of which have their origins in EPC.  As a rather general problem in many condensed matter systems, the same issue has  arisen not only in other transition metal dichalcogenides in both bulk and monolayer forms\citep*{nagata1992superconductivity,freitas2016strong,valla2000charge}, but also in high-temperature superconductors, where the competition between superconductivity and CDW has been noted~\citep*{Ghiringhelli2012,LeTacon2013,Chang2012}. Combined with its relatively simple crystal structure,  monolayer NbSe$_2$ provides an ideal platform to the study of the competition between superconductivity and CDW based on first principles calculations, since the EPC in this material can be studied in great detail.  In  previous work\cite{zheng2017charge,Lian2018}, the competition for the density of states (DOS) at Fermi energy between superconductivity and CDW in monolayer NbSe$_{2}$, and, in particular, the opening of bandgap on parts of the Fermi surface after CDW  were noted~\cite{zheng2017charge}. But the microscopic mechanism regarding how they compete remains unclear. The superconducting gap structure in monolayer limit has yet to be analyzed. For example, two-gap superconductivity is observed in bulk NbSe$_{2}$~\citep{Noat2015}, but not in thin-film samples~\citep{Khestanova2018}, which  clearly requires clarification. Moreover, there remain uncertainties regarding the superconducting transition temperatures ($T_{c}$) experimentally\citep{Xi2015,ugeda2016,Xing2017} in this material, possibly due to the fact that a 2-dimensional system is easily affected by the factors  extrinsic to the material. An essential step towards understanding  the uncertainty necessarily starts with a quantitative  understanding of the superconductivity in pristine (weakly coupled with substrate, no doping) monolayer NbSe$_2$.

In this work, we show that the superconductivity and CDW are responsible for the formation of gaps in distinct sectors of the Fermi surface, associated with intra-pocket and inter-pocket scatterings, respectively. The material becomes fully gapped only when superconductivity and CDW coexist. Furthermore, we show the  absence of  two-gap feature in monolayer limit is possibly due to extensive CDW gap on the Fermi surface in $\mathbf{K}$ and $\mathbf{K'}$ Fermi pockets. Our study also suggests that the factors extrinsic to the material have different influences on the superconductivity. An overall blue shifts of the phonons and sharp decrease of the  Eliashberg spectrum are found. 
  
 Density-functional theory and density-functional perturbation theory  calculations are performed to investigate the crystal structures, electronic structures, phonons, EPC and superconducting properties of monolayer NbSe$_2$ in non-CDW and CDW phases\citep{Baroni1987,Giannozzi2009,Perdew1996}. A Brillouin zone unfolding scheme is devised~\cite{Allen2013}(see Sec. S8 in supplement) to unfold the Fermi surface, phonon dispersion and superconducting gap of the CDW  phase from supercell Brillouin zone to the primitive one. The calculation of Eliashberg spectra [$\alpha^{2}F(\omega)$] are first performed on coarse $\mathbf{k}$- and  $\mathbf{q}$ grids and then interpolated onto denser grids ~\cite{Ponce2016,mostofi2008wannier90}. $\mathbf{k}$ and $\mathbf{q}$ indicate the quasi-momenta of electrons and phonons, respectively. We use the notion $\text{P}_{i}$ ($i = \mathbf{\Gamma},~\mathbf{K},~\mathbf{K'}$) to indicate the Fermi pockets $\mathbf{\Gamma},~\mathbf{K},~\mathbf{K'}$, respectively. Fermi-pocket resolved  matrix elements $g_{\mathbf{k}\in{\text{P}_{i}},\mathbf{k'}\in\text{P}_{j}}^{\nu}$ are analyzed, which quantify  scattering amplitude via  phonons with $\mathbf{q} = \mathbf{k'} - \mathbf{k}$  and branch $\nu$ between $\mathbf{k}$ and $\mathbf{k'}$  from  $\text{P}_{i}$ and $\text{P}_{j}$, respectively. Total EPC constant $\lambda$ reads: $\lambda=2\int\alpha^2F(\omega)/\omega d\omega$, where $\lambda_{\mathbf{q}\nu}$ is an EPC constant associated with $\mathbf{q}$ and branch $\nu$. The temperature dependent superconducting gaps are calculated~\cite{mostofi2008wannier90,Ponce2016}.  The  Allen-Dynes-modified McMillan equation~\citep*{Allen1975,McMillan1968,Giustino2017} is also employed to  evaluate $T_{c}$, where  an effective Coulomb potential $\mu_{\text{c}}^{*}$ is set to  be 0.15,  estimated according to the DOS at Fermi energy in the CDW phase~\cite{zheng2017charge}: $\mu_{\text{c}}^{*}\approx [0.26N_{\text{F}}/(1+N_{\text{F}})]$ \citep*{Garland1972,Xi2016}. More  computational details are shown in the Sec.~S2 of supplement.
 %
\begin{figure}
	\centering
	\includegraphics[width=76 mm]{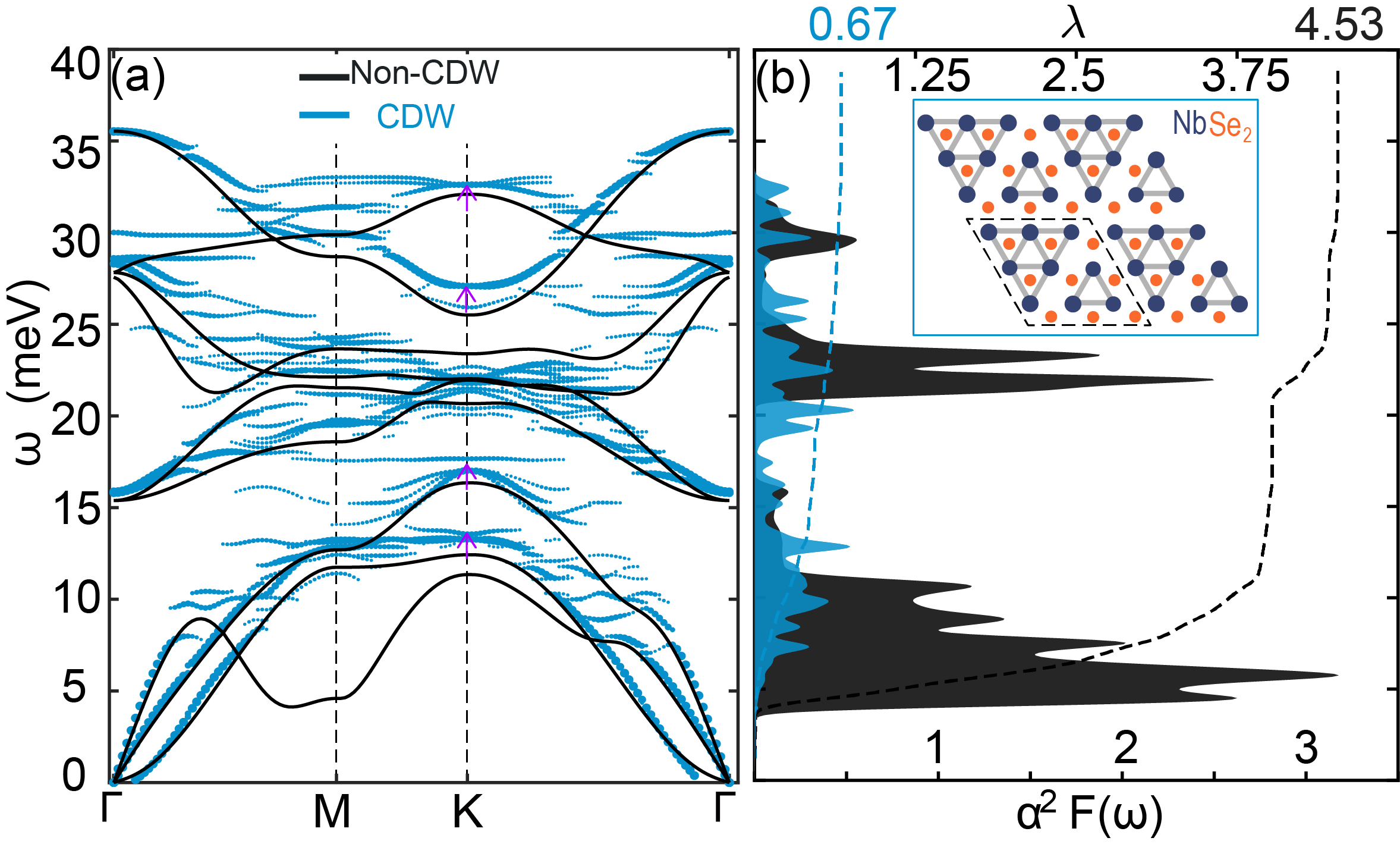} \\	
	\caption{ (a) unfolded phonon dispersion of the CDW phase (light-blue dots, $\sigma = 0.01$ Ry) in primitive Brillouin zone, combined with the dispersion of the non-CDW (black curves, 0.03 Ry). The sizes of the  dots are positive correlated to the unfolding weights. Those $\omega_{\mathbf{q}}$ with the weights $<$ 1/9 are not shown. The pink arrows label the blue shifts of the phonons. (b)Eliashberg spectra $\alpha^2F(\omega)$ with integrated EPC  constants $\lambda(\omega)$ (colored dashed lines) of the non-CDW and CDW phases. The inset is a top view of the CDW phase. The solid gray bonds between the adjacent Nb atoms indicate their distances are shortened after CDW transition. The rhombus bounded by dashed lines donates a 3$\times$3 CDW supercell.}
	\label{fig1:str_pho}
\end{figure} 



 The structure of non-CDW monolayer NbSe$_{2}$ forms a 2-dimensional hexagonal lattice, which is composed of 3 layers of atoms, with a Nb layer sandwiched between Se layers. Each Nb atom sits inside a trigonal prismatic cage formed by 6 nearest-neighbor Se atoms. Nb atoms form a perfect hexagonal closest packing  structure with the  shortest  Nb-Nb separation  $a = 3.48$ \AA~according to our calculation. The phonon dispersions of the non-CDW phase are computed and displayed in Fig.~S1(b). The pink curves are computed using a regular electronic broadening ($\sigma  = 0.01 $ Ry) in the self-consistent calculation of charge density. The lowest phonon branch exhibits the minimum imaginary frequency at about 2/3$\mathbf{\Gamma}\mathbf{M}$ ($\mathbf{q}_{\textrm{CDW}} = \frac{1}{3}\mathbf{a}^{*}$), indicating  CDW instability corresponding to a $3\times3$ supercell in real space~\cite{ugeda2016,xi2015ising,zheng2017charge}. It should be mentioned that two reconstructed and dynamically stable structures in the CDW supercell have been reported\cite{silva2016electronic,zheng2017charge,Lian2018} theoretically, one of which was suggested to be the ground-state structure (named ``triangle'' phase in~\cite{zheng2017charge}),  which is recently confirmed by the calculation of Raman active modes (``3+6-hollow'' in ~\citep*{Lian2018}). Therefore, our calculations are mainly performed on the ground-state structure, which is shown in the inset of Fig.~\ref{fig1:str_pho}(b). For completeness, the electron-phonon coupling of the remaining structure is also qualitatively analyzed [Sec.~10 in the supplement].

A primary objective of this work is to compare the EPC between the non-CDW and CDW phases. The former is problematic computationally due to the imaginary phonons around $\mathbf{q}_{\text{CDW}}$ which is usually excluded in EPC calculation. We adapt a larger $\sigma = 0.03$ Ry to stabilize the non-CDW crystal so that the CDW related phonons can be involved in EPC calculation. It is found that the large $\sigma$ removes the imaginary phonons, and the CDW instability is preserved as dip phonons around $\mathbf{q}_{\text{CDW}}$ [black curves in Figs.~\ref{fig1:str_pho}(a) and S1(b)].  Except for the above observation, there is no other significant change in the phonon spectrum. Due to the CDW phonons are sensitively related to electronic broadening (temperature), our results of EPC calculation for the non-CDW phase only serve as a qualitatively analysis. The details of the above discussion are shown in Sec.~S1 of the supplement.

 

To investigate the influence of CDW on the phonons, the phonon dispersion of the CDW phase is computed  (using $\sigma = 0.01$ Ry) and unfolded to the primitive Brillouin zone. The result is displayed in Fig.~\ref{fig1:str_pho}(a)  together with the dispersion of the non-CDW phase. The removal of the imaginary phonons around $\mathbf{q}_{\textrm{CDW}}$ indicates the stability of the CDW phase. Furthermore, it is found that phonons of the CDW  phase display  an overall blue shift compared to the non-CDW phase. As the CDW transition is largely driven by Nb net and induces the formation of the triangle pattern, it is expected that the blue shift is derived from the Nb lattice. This is indeed the case which can be seen from Figs.~\ref{fig1:str_pho}(a) and S3(c), the blue-shifted phonon density of states is dominated by Nb (three acoustic and the two highest phonon branches). Part of the blue-shifted phonons at $\mathbf{K}$ is marked using pink arrows as displayed in Fig.~\ref{fig1:str_pho}(a).

To understand the superconductivity of  monolayer NbSe$_2$, $\alpha^2F(\omega)$ of the CDW phase is calculated and displayed in Fig.~\ref{fig1:str_pho}(b) (light-blue region) along with that of  the non-CDW  phase (black region). The most remarkable change in $\alpha^2F(\omega)$ of the CDW phase is the sharp reduction of its intensity, which is due to  the formation of bandgap induced by CDW as we analyzed in Sec.~S4 of the supplement. As a result, the resultant $\lambda$  decreases from 4.53 (non-CDW phase) to 0.67 after CDW [Fig.~\ref{fig1:str_pho}(b)]. Using the Allen-Dynes-modified McMillan equation with $\mu_{\text{c}}^{*}=0.15$, $T_{\text{c}}$ is estimated to be 2.7 $K$, with $\omega_{log} $ = 144.5 $K$. The computed $T_{c}$ here agree well with experiments~\citep{ugeda2016,Xi2015}, though a bit lower than the recent theoretical work (4.5 $K$)~\citep{Lian2018}, which is directly due to the smaller computed $\lambda$  here (0.67 vs 0.84), possibly owning to the Wannier interpolation used in this work.

To evaluate the superconducting gap function $\Delta_{\mathbf{k}}$ in the CDW phase, the anisotropic Migdal-Eliashberg formalism are employed~\cite{Ponce2016}. Histograms of $\Delta_{\mathbf{k}}$ on and near the  Fermi surface  at different temperatures are displayed in Fig.~\ref{fig2:sc}(a).  When $T$ = 2 $K$, the values of $\Delta_{\mathbf{k}}$ are within the range between 0.5 and 0.8 meV. In particular, a pronounced peak centered at about 0.65 meV is observed, indicating a highly concentrated distribution of $\Delta_{\mathbf{k}}$ at that energy. As $T$ increases from 2.0 to 4.4 $K$, the range of  $\Delta_{\mathbf{k}}$ become narrower, with the gradual decrease of the energy and height  of  the peak, which finally disappears at about 4.4 K. Therefore, $T_{\text{c}}$ can be estimated to be 4.4 $K$ accordingly, which falls in the range between the experimental results~\citep{Xi2015,Xing2017}. The computed superconducting DOS  also exhibits consistent trend (Sec.~S5 in the supplement). The computed sizes of $\Delta_{\mathbf{k}}$ are consistent with a recent experimental work~\citep{Khestanova2018}, where the gap is probed to be 0.6 meV at 0.3 $K$ for bilayer NbSe$_2$. The resultant $T_{c}$ using the Allen-Dynes-modified McMillan equation  (2.7 $K$) and anisotropic Migdal-Eliashberg formalism (4.4 $K$) are different.  The latter result should  be more reliable for monolayer NbSe$_2$, since it is more suitable for the layered system of reduced dimensionality, where anisotropic of the Fermi surface is important~\cite{Margine2013}. The computed $T_{c}\approx 4.4~K$ for the pristine monolayer NbSe$_2$ here is found to be slightly different from either of the experimental results~\cite{Xi2015,ugeda2016,Xing2017}, but lie in the ranges of them. Considering the fact that the samples in these experiments were prepared using different methods on substrates and measured with/without capping layer, our study suggests that the above extrinsic factors should have different influences on the superconductivity. 

 To further understand the superconducting property microscopically, $\Delta_{\mathbf{k}}(\omega=0, T=2K)$ of the CDW phase is computed and unfolded to primitive Brillouin zone [Fig.~\ref{fig2:sc}(c)]. Fig.~\ref{fig2:sc}(b) shows that the Fermi surface is partially or fully gapped when the CDW is formed~\citep{zheng2017charge}, which is in qualitatively agreement with experimental observation that a striking dip at Fermi energy (but does not reach to zero) was seen in the dI/dV spectrum~\citep*{ugeda2016}. The experimental striking dip is mainly due to the extensive Fermi surface gapping in $\mathbf{K}$ and $\mathbf{K'}$ pockets according to the computed Fermi surface~[Fig.~\ref{fig2:sc}(b)]. Furthermore, it is found that the remnant of the Fermi surface after CDW [Fig.~\ref{fig2:sc}(c)] will be fully gapped in the coexistent phase with almost constant value of  $\Delta_{\mathbf{k}}\approx0.65$ meV, consistent to the remarkable peak center at 0.65 meV [Fig.~\ref{fig2:sc}(a)]. Besides, a fraction of $\mathbf{k}$ in $\mathbf{\Gamma}$ pocket exhibit smaller $\Delta_{\mathbf{k}}\approx0.5$ meV [blue sectors in Fig.~\ref{fig2:sc}(c)]. Meanwhile, a fraction of $\mathbf{k}$ with larger $\Delta_{\mathbf{k}}\approx0.8$ meV in $\mathbf{K}$ and $\mathbf{K'}$ pockets is seen [yellow sectors in Fig.~\ref{fig2:sc}(c)]. The distribution of $\Delta_{\mathbf{k}}$ here is similar to the case of NbS$_2$, where two small and a large peaks of $\Delta_{\mathbf{k}}$ are found to stem from $\mathbf{\Gamma}$ and  $\mathbf{K}$, $\mathbf{K'}$ Fermi sheets, respectively [Fig.~5(a) in~\cite{Heil2017}].

 We now show that the absence of two-gap feature in few-layer NbSe$_{2}$ experimentally~\citep{Khestanova2018} is possibly due to the extensive  Fermi surface gapping in $\mathbf{K}$ and  $\mathbf{K'}$ pockets induced by CDW. The two-gap feature is found in the non-CDW phase with 3 intact Fermi pockets and we further assign the large gap to $\mathbf{K}$ and  $\mathbf{K'}$ pockets dominated by the in-plane orbitals of Nb, the small gap to the $\mathbf{\Gamma}$ pocket (out-of-plane orbitals of Nb). The assignment of the gaps above is similar to its bulk counterpart~\cite{Noat2015} and a related material, NbS$_{2}$~\cite{Heil2017}.  More details can be found in Sec.~S6 of the supplement. Therefore, the absence of the two-gap feature in the monolayer computationally~[Fig.~\ref{fig2:sc}(a)]  is possibly due to the extensive bandgap of  the $\mathbf{K}$  and $\mathbf{K'}$ pockets [Fig.~\ref{fig2:sc}(b)], removing the states that will exhibit large $\Delta_{\mathbf{k}}$ if they were not gapped by CDW. This is also corroborated by our result that those $\mathbf{k}$, located near the boundary of partial and fully CDW gapped sectors of  $\mathbf{K}$  and $\mathbf{K'}$ pockets [yellow sectors in Fig.~\ref{fig2:sc}(c)], exhibit relatively large  $\Delta_{\mathbf{k}}$. The absence of two-gap feature here is consistent with recent experimental observation, where such feature is not seen in few-layer NbSe$_2$~\citep{Khestanova2018}.

\begin{figure}
	\centering
	\includegraphics[width=85mm]{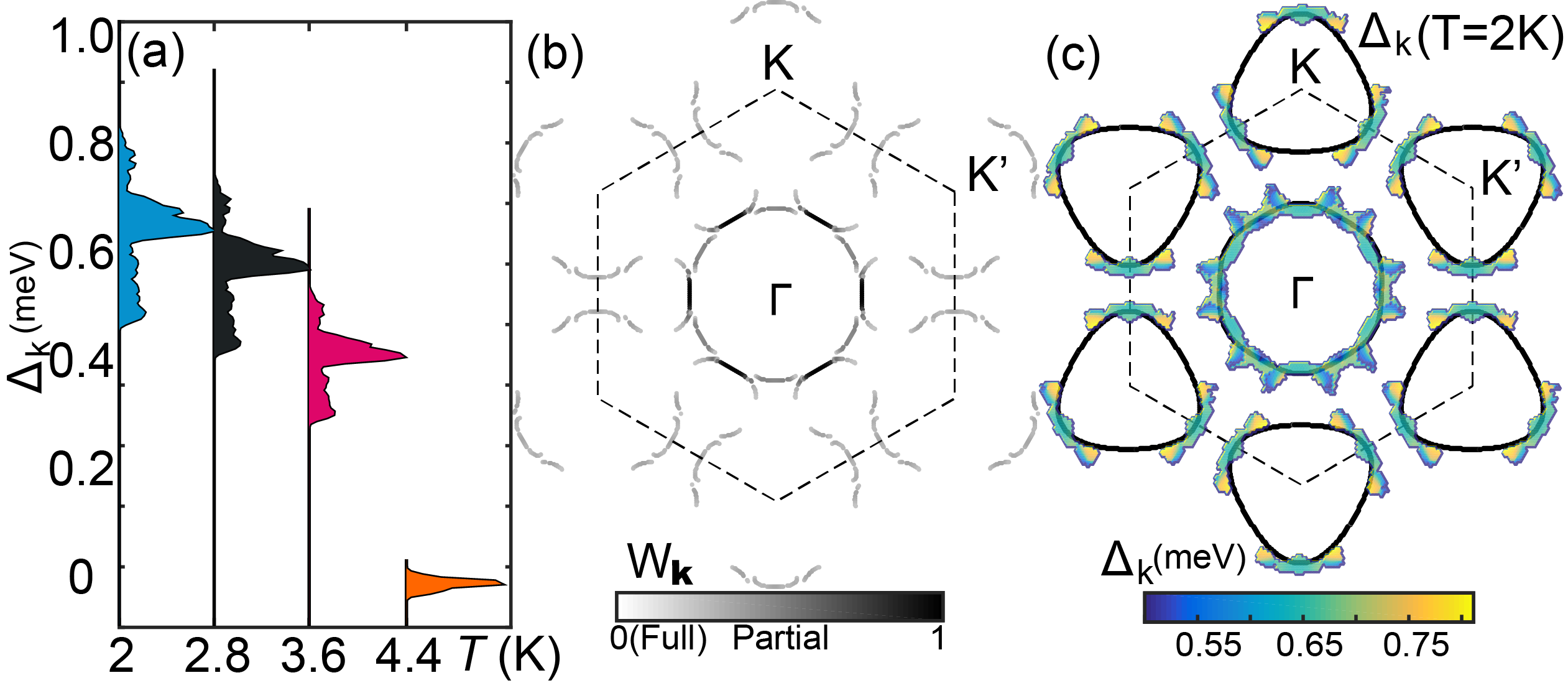} \\
	\caption{ (a) Computed anisotropic superconducting gap $\Delta_{\mathbf{k}}$ of the CDW phase around Fermi surface, evaluated as a function of temperature. For each temperature, the histograms are related to  the number of states on the Fermi surface with that superconducting gap energy. (b) Unfolded Fermi surface of the CDW phase. The gray level of the Fermi sectors donate the corresponding unfolding weights $W_{\mathbf{k}}$. The colorbar of $W_{\mathbf{k}}$ is shown, indicating the partial or full bandgap of $\mathbf{k}$ in the CDW phase. (c) The black circles donate the unfolded Fermi surface of the non-CDW  phase using a $3\times3$ supercell. The colored sectors represent the dominant $\Delta_{\mathbf{k}}(T=2 K)$ of the CDW phase in primitive Brillouin zone obtained by the unfolding scheme. }
	\label{fig2:sc}
\end{figure}


\begin{figure}
	\centering
	\includegraphics[width=76mm]{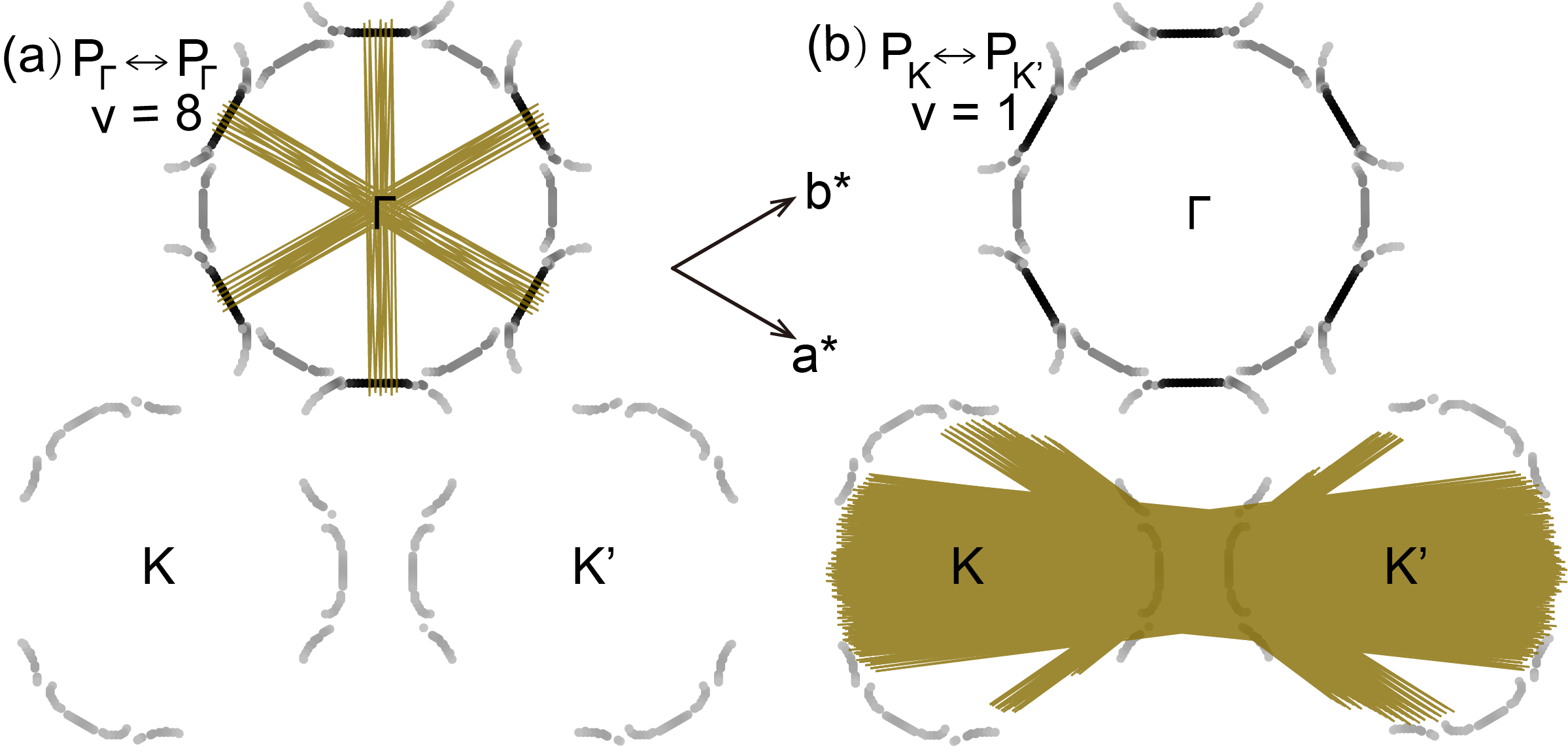} \\
	\caption{ 
    Visualization of  pocket-resolved matrix elements $\left |g_{\mathbf{k} \in\text{P}_{\mathbf{\Gamma}},\mathbf{k'} \in\text{P}_{\mathbf{\Gamma}}}^{\nu=8} \right |$ (a), $\left |g_{\mathbf{k} \in\text{P}_{\mathbf{K}},\mathbf{\mathbf{K'}} \in\text{P}_{j}}^{\nu=1} \right |$ (b). Those elements with dominant values for the above two-type matrix elements are first picked out. Each pair of $\mathbf{k}$, $\mathbf{k'}$ related to the selected matrix elements  is subsequently located and visualized by drawing the connecting vector $\mathbf{q} = \mathbf{k'} - \mathbf{k}$ using green line as shown in (a) and (b). The black and gray lines in (a) and (b) indicate the unfolding Fermi surface of the CDW phase, which is the same as Fig.~\ref{fig2:sc}(b).}
	\label{fig4:kk_connec}
\end{figure}

We have shown that the remnant Fermi surface after CDW is associated with superconductivity. A natural question is how the superconducting gaps are formed. We then show that they are formed possibly due to the intra-pocket scatterings. EPC are decomposed based on the topology of the Fermi surface: the intra-pocket scattering within the $\mathbf{\Gamma}$ pocket ($\text{P}_{\mathbf{\Gamma}}\leftrightarrow\text{P}_{\mathbf{\Gamma}}$),  $\mathbf{K}$  and $\mathbf{K'}$ pockets ($\text{P}_{\mathbf{K}}\leftrightarrow\text{P}_{\mathbf{K}}$), the inter-pocket scattering between the $\mathbf{K}$  and $\mathbf{K'}$ pockets ($\text{P}_{\mathbf{K}}\leftrightarrow\text{P}_{\mathbf{K'}}$), and  between the $\mathbf{\Gamma}$ and $\mathbf{K}$ , $\mathbf{K'}$ pockets ($\text{P}_{\mathbf{\Gamma}}\leftrightarrow\text{P}_{\mathbf{K}}$). Pocket-resolved EPC matrix elements $g_{\mathbf{k}\in\text{P}_{i},\mathbf{k'}\in\text{P}_{j}}^{\nu}$ are  analyzed. The distributions of $\left|g_{\mathbf{k}\in\text{P}_{i},\mathbf{k'}\in\text{P}_{j}}^{\nu} \right|$ are displayed in Fig.~S3(g). By visualizing the pocket-resolved EPC matrix elements as described in the caption of Fig.~\ref{fig4:kk_connec}, we find that those $\mathbf{k}$ with the largest unfolding weights in the $\mathbf{\Gamma}$ pocket (black sectors), where  $\Delta_{\mathbf{k}}\approx 0.65$ meV [Fig.~\ref{fig2:sc}(c)], are in favor of mutual scattering via optical phonons ($\nu=8$) [Fig.~\ref{fig4:kk_connec}(a)]. By similar analysis as shown in Figs.~S9(a-b)  of the supplement, we find that the formation of the remaining $\Delta_{\mathbf{k}}$ on $\mathbf{\Gamma}$ pockets are also related to the intra-pocket scattering within $\mathbf{\Gamma}$ pockets. The above findings indicate that the intra-pocket scatterings  are related to the superconductivity in this material.

\begin{figure}
	\centering
	\includegraphics[width=76mm]{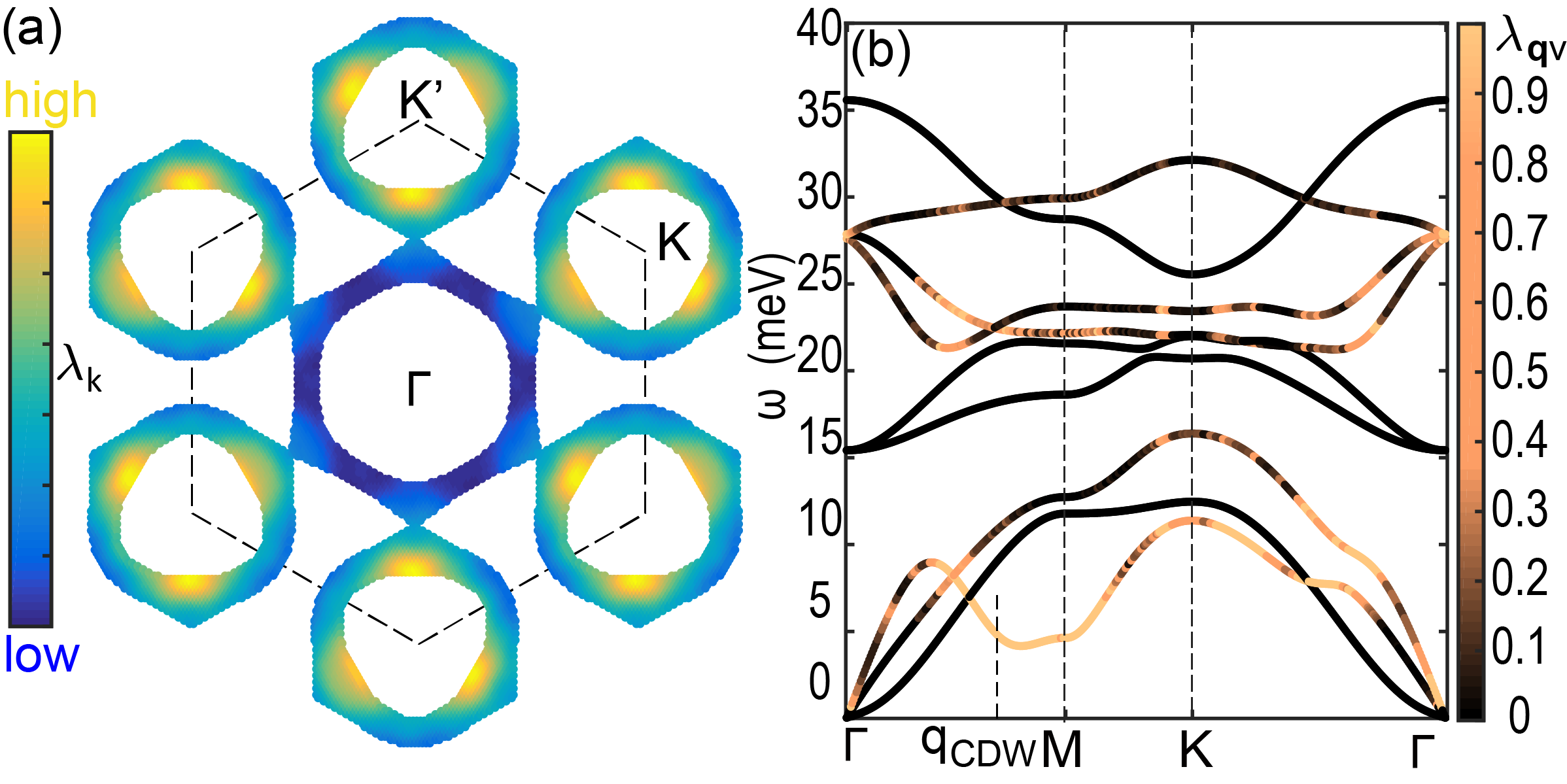} \\
	\caption{ (a) Distribution of  $\lambda_{\mathbf{k}}$ of the non-CDW around the Fermi surface. The black dashed lines represent the boundary of the Brillouin zone. (b) Phonon dispersion, $\omega_{\mathbf{q}\nu}$, of the non-CDW phase. The color code depicts the dimensionless EPC parameter $\lambda_{\mathbf{q}\nu}$. All the position of $\omega_{\mathbf{q}\nu}$ where  $\lambda_{\mathbf{q}\nu} > 1$  are colored with the same color for easy visualization.  A similar plot using a different colorbar can be found in Fig.~S2 in supplement.  
	}
	\label{fig3:eph1}
\end{figure}

We further assign the formation of the bandgaps induced by CDW  to inter-pocket scatterings. It is found that the extensive of Fermi surface gapping induced by CDW [Fig.~\ref{fig2:sc}(b)] at a given $\mathbf{k}$ is positively correlated to the value of $\lambda_{\mathbf{k}}$ as shown in Fig.~\ref{fig3:eph1}(a). In particular,  those $\lambda_{\mathbf{k}}$, whose $\mathbf{k}$ come from the fully band-gapped sectors in $\mathbf{K}$ and $\mathbf{K'}$ pockets, are significant as shown by yellow in Fig.~\ref{fig3:eph1}(a), indicating that they couple strongly with phonons. By further locating the positions of $\mathbf{k}$ and $\mathbf{k'}$ for the dominant $\left|g_{\mathbf{k}\in\text{P}_{\mathbf{K}},\mathbf{k'}\in\text{P}_{\mathbf{K'}}}^{\nu=1} \right|$ as shown in Fig.~\ref{fig4:kk_connec}(b), we find that these $\mathbf{k}$ and $\mathbf{k'}$ are mainly located on the fully band-gapped sectors. Meanwhile, each pair of the above $\mathbf{k}$ and $\mathbf{k'}$ is connected by the phonons with momentum $\mathbf{q} = \mathbf{k'} - \mathbf{k}$, which is found to be around $\mathbf{q_{\text{CDW}}}$.  Consistently, the above phonons exhibit strong coupling with electrons, as the corresponding $\lambda_{\mathbf{q},\nu=1}$ are significant [see Figs.~\ref{fig3:eph1}(b) and S8], which are directly associated with CDW instability. It is also found that the electronic scattering between $\mathbf{\Gamma}$ and $\mathbf{K}$, $\mathbf{K'}$ pockets  is related with the CDW as shown in Fig.~S9(c) in the supplement. The above findings indicate that the inter-pocket scatterings is related to the CDW in this material. In particular, the scattering between $\mathbf{K}$ and $\mathbf{K'}$ pockets via acoustic phonons with $\mathbf{q} \approx \mathbf{q}_{\text{CDW}}$ should be the driving force. The above results are consistent with previous knowledge that the CDW in this material is driven by momentum-resolved EPC matrix elements~\cite{Johannes2008,Zhu2015}.


In summary, the competition and coexistence of the superconductivity and  CDW in pristine monolayer NbSe$_2$ has been scrutinized using first principles calculations, leading to microscopic understanding of the interplay among these orders. First, the mechanism of the competition  between the superconductivity and CDW has been quantitatively analyzed. In the non-CDW phase,  three intact Fermi pockets centered at $\mathbf{\Gamma}$, $\mathbf{K}$ and $\mathbf{K'}$ lead to $\lambda$ = 4.53 and $T_{c} \approx 18~K$ [see Fig.~S5(a)], which is higher than experimental values.  Once the CDW is formed, bandgaps form at parts of the Fermi surface, where EPC is large in the non-CDW phase. These partial or full gaps lead to the reduction of computed $\lambda =~0.67$ and $T_{c} \approx~4.4 K$, which  are within the range of experimental results. The distribution of $\Delta_{\mathbf{k}}$ on the remnant of Fermi surface in the CDW phase is also visualized. Combined with the analysis of EPC matrix, we assign the intra-pocket scatterings of electrons to the formation of superconducting gaps, and the inter-pocket scatterings to the bandgaps induced by CDW. Interestingly, the absence of the two-gap feature in the superconducting monolayer is found to be due to the extensive CDW gap in $\mathbf{K}$ and $\mathbf{K'}$ pockets. From our calculations, $T_{c}$ in the CDW phase is estimated to be $4.4~K$, which lies in the ranges of experimental results, but  different from them. This  suggests that the environmental or non-intrinsic factors, such as substrates and capping layers, have different influences on the superconductivity in monolayer NbSe$_{2}$. Finally, overall blue shifts of phonons are observed in the CDW phase compared to that of the non-CDW phase, giving rise to some prominent changes in $\alpha^2F(\omega)$.

\begin{acknowledgements} 
This work is supported by Ministry of Science and Technology of the People's Republic of China (Grant Nos. 2018YFA0305601, 2016YFA0301004), National Natural Science Foundation of China (Grant Nos. 11725415 and 11804118), and by Strategic Priority Research Program of Chinese Academy of Sciences, Grant No. XDB28000000. Part of the calculations were performed on the Tianhe-I Supercomputer System. 		
\end{acknowledgements}

\bibliographystyle{apsrev4-1}
%



\setcounter{figure}{0}
\setcounter{page}{0}
\renewcommand{\thepage}{S\arabic{page}}  
\renewcommand{\thesection}{S\arabic{section}}   
\renewcommand{\thetable}{S\arabic{table}}   
\renewcommand{\thefigure}{S\arabic{figure}}
\renewcommand{\theequation}{S\arabic{equation}}

\onecolumngrid

\newpage

\title{\textit{Supplemental Material}:\\Electron-phonon Coupling and the Coexistence of Superconductivity and Charge-Density Wave  in Monolayer NbSe$_2$}

\author{Feipeng Zheng}
\email{fpzheng_phy@163.com}
\affiliation{Siyuan Laboratory, Guangzhou Key Laboratory of Vacuum Coating Technologies and New Energy Materials, Department of Physics, Jinan University, Guangzhou 510632, China}

\author{Ji Feng}
\email{jfeng11@pku.edu.cn}
\affiliation{International Center for Quantum Materials, School of Physics, Peking
	University, Beijing 100871, China}
\affiliation{Collaborative Innovation Center of Quantum Matter, Beijing 100871,
	China}
\affiliation{CAS Center for Excellence in Topological Quantum Computation, University of Chinese Academy of Sciences, Beijing 100190, China}

\maketitle

\section{The calculations of non-CDW phase using the large electronic broadening}

\begin{figure}[H]
	\centering
	\includegraphics[width=140mm]{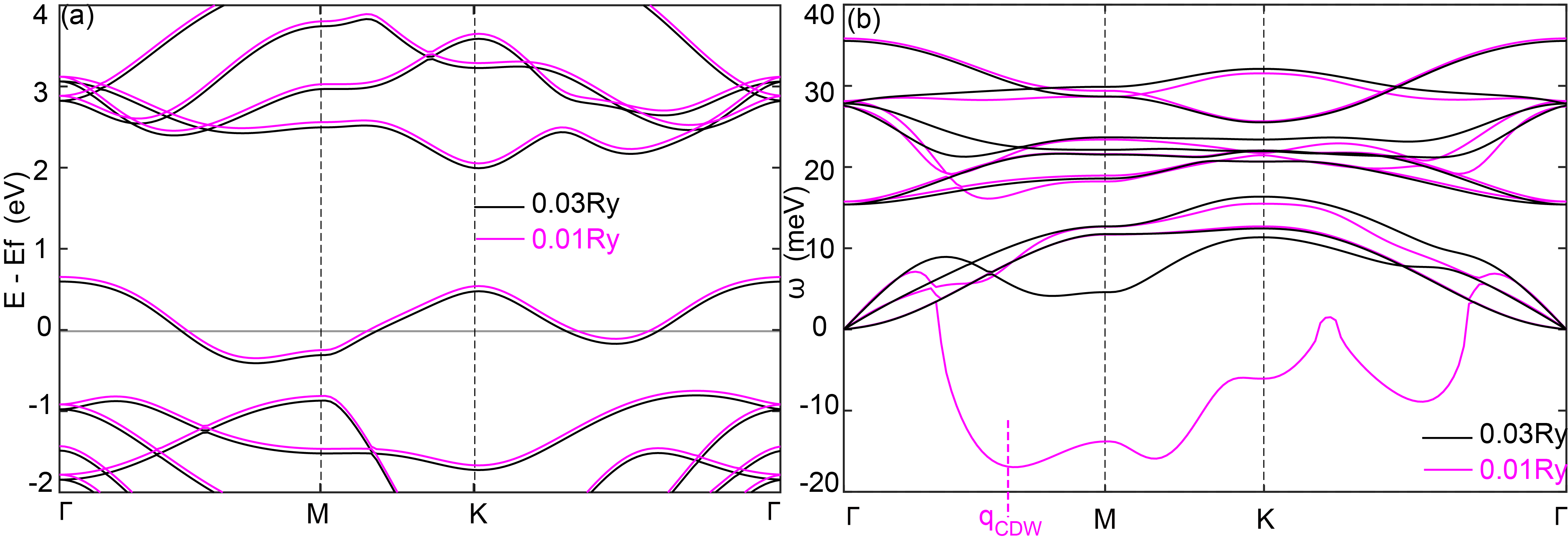} \\
	\caption{ Computed electronic bandstructure (a) and Phonon dispersions (b) of the non-CDW phase using a regular (0.01 Ry, pink curves) and a large (0.03 Ry, black curves) electronic broadening during the calculation of the charge-density self-consistently.  
	}
	\label{figs1:phonon}
\end{figure}

An essential purpose of our work is to  investigate how electron-phonon coupling come into play in the formation of CDW in monolayer NbSe$_{2}$, by comparing related physical quantities (such as Eliashberg spectra) of the systems before and after CDW transition. The latter one can be obtained by directly computing electron-phonon coupling of the CDW phase, during which a regular electronic broadening is used to calculate its charge density self-consistently.  The electron-phonon coupling before CDW transition can be obtained by computing  the non-CDW phase similarly at first glance. However, such a calculation of the non-CDW phase lead to the imaginary phonons around $\mathbf{q}_{\text{CDW}}$ [Fig.~\ref{figs1:phonon}(b), pink curves], which are frequently excluded in electron-phonon coupling calculation by the mainstream package (such as EPW package used in this work), due to the ill-definition of electron-phonon coupling of imaginary phonons.

To tackle with this problem, we adapt a large electronic broadening (0.03 Ry) in the self-consistent calculation of the charge-density of the non-CDW phase, so that the crystal is stable and the CDW instability is also preserved in terms of dip phonons around $\mathbf{q_{\text{CDW}}}$ [black curves in Fig.~\ref{figs1:phonon}(b)]. As a result, the CDW related phonons can be taken into account in the electron-phonon coupling calculations, and the influence of the CDW can now be analyzed. This is reasonable, since the non-CDW structure is indeed stable when $T > T_{\text{CDW}}$, and using a large broadening can be understood as the increase of electronic temperature, which stable the non-CDW crystal.  We carefully examine the relaxed crystal structure of the non-CDW phases using the broadening of 0.01 and 0.03 Ry. We find they are almost identical (the relaxed lattice constants in the above two cases are 3.483 and 3.485~\AA,respectively). We also find that electronic bandstructures computed using the two broadening are similar as shown in Fig.~\ref{figs1:phonon}(a). Only a tiny shift of Fermi energy is observed due to the different electronic occupations in the electronic broadenings of 0.03 and 0.01 Ry. For the phonon dispersion, the large broadening pulls the imaginary phonons to positive [Fig.~\ref{figs1:phonon}(b)]. Except this, there is no significant changes. In fact, a similar method was used  to investigate the influence of electron-phonon coupling on electronic transport properties\cite{Hinsche2017}.

To sum up, the purpose of using a large electronic broadening of 0.03 Ry to calculate the charge density of the non-CDW phase is to prevent the non-CDW crystal from CDW transition, i.e. pull the imaginary phonons around $\mathbf{q}_{\text{CDW}}$ to positive, so that the CDW related phonons can be taken into account in the electron-phonon coupling calculations, and the influence of the CDW can be analyzed.

\section{Computational details}

We use Norm-conserving pseudopotentials \citep{hamann1979norm} to describe the interactions between valence electrons and ionic cores of both Nb and Se atoms. The Kohn-Sham valence states are expanded in the plane wave basis set with a kinetic energy truncation at 90 Ry. An 18$\times$18$\times$1 and a 6$\times$6$\times$1  $\mathbf{k}$-grid centered at $\mathbf{\Gamma}$  are chosen for the  calculations of the charge density of ground states for the non-CDW and CDW phases respectively in combination with Marzari-Vanderbilt-type  broadening~\cite{Marzari1999} of 0.01 and 0.03 Ry. The equilibrium crystal structures are determined by a conjugated-gradient relaxation of ionic positions until the Hellmann-Feynman force on each atom is less than 0.001 \AA /eV and  zero-stress tensors are obtained. The phonon spectra of the two phases are computed. A 9$\times$9$\times$1 and a 3$\times$3$\times$1 $\mathbf{q}$-grid centered at $\mathbf{\Gamma}$  are chosen for the sampling of phonon momenta for the non-CDW and the CDW phases, respectively. To directly compare the phonons of the CDW phase to that of the non-CDW, a Brillouin zone unfolding scheme is devised~\cite{Allen2013}(see Sec.~S8 in the supplement). Electron-phonon coupling matrix elements of the non-CDW  and CDW phases are first computed on the above coarse $\mathbf{k}$- and $\mathbf{q}$-grid~\cite{Ponce2016,mostofi2008wannier90}. Then, they are interpolated~\cite{mostofi2008wannier90} to denser grids, i.e. a 120$\times$120$\times$1 $\mathbf{k}$- and $\mathbf{q}$-grid, respectively, for the non-CDW phase, and a 40$\times$40$\times$1 $\mathbf{k}$- and $\mathbf{q}$-grid for the CDW phase. The convergence of $\alpha^{2}F(\omega)$ with respect to Brillouin zone sampling is confirmed as shown in Fig.~\ref{figs6:a2f_conv}(a) in the supplement. After that, the electron-phonon coupling constants and Eliashberg spectra $\alpha^{2}F(\omega)$ ~\cite{Grimvall1981} are computed. 
The Fermi-pocket-resolved (henceforth pocket-resolved) Eliashberg spectrum $\alpha^2F(\text{P}_{i}\leftrightarrow\text{P}_{j},\omega)$ can be  obtained by restricting  the integrations of electronic momenta on specific Fermi pockets in the formula of $\alpha^{2}F(\omega)$:
\begin{multline}
	\alpha^{2}F(\text{P}_{i}\leftrightarrow\text{P}_{j},\omega) = \frac{S^2}{N_{F}}\sum_{\nu}\int_{\mathbf{k}\in\text{P}_i} \frac{d^2\mathbf{k}}{(2\pi)^2}\int_{\mathbf{k'}\in\text{P}_j}\frac{d^2\mathbf{k'}}{(2\pi)^2} \times\left|g_{\mathbf{k},\mathbf{k'}}^{\nu} \right|^{2}\delta(\epsilon_{\mathbf{k}} - \epsilon_{\text{F}})\delta(\epsilon_{\mathbf{k'}} - \epsilon_{\text{F}})\delta(\omega - \omega_{\mathbf{k'}-\mathbf{k},\nu}),
	\label{eq:a2f}
\end{multline}
where $\text{P}_{i}$ or $\text{P}_{j}$ indicate  Fermi pocket $i$ or $j$, respectively. Since there are three distinct Fermi pockets centered at $\mathbf{\Gamma}$, $\mathbf{K}$ and $\mathbf{K'}$ respectively for the non-CDW phase, $i, j = \mathbf{\Gamma}, \mathbf{K}, \mathbf{K'}$ here. $N_{\text{F}}$ is the density of state (DOS) at Fermi energy $\epsilon_{\text{F}}$, $S$ the area of the unit cell,  $\omega_{\mathbf{k'}-\mathbf{k},\nu}$ the frequency of a phonon with momentum $\mathbf{k'} - \mathbf{k}$ and a branch $\nu$. $g_{\mathbf{k},\mathbf{k'}}^{\nu}$ ($\mathbf{k} \in \text{P}_{i}$ and $\mathbf{k'} \in \text{P}_{j}$) or $g_{\mathbf{k}\in{\text{P}_{i}},\mathbf{k'}\in\text{P}_{j}}^{\nu}$,  is named Fermi-pocket resolved electron-phonon coupling matrix element here, which quantify  scattering amplitude via  phonons with  momentum $\mathbf{q} = \mathbf{k'} - \mathbf{k}$  and branch $\nu$ between $\mathbf{k}$ and $\mathbf{k'}$  from  Fermi pockets $\text{P}_{i}$ and $\text{P}_{j}$ respectively. The total electron-phonon coupling constant $\lambda$ reads: 
\begin{equation}
	\lambda=\sum_{\mathbf{q}\nu}\lambda_{\mathbf{q}\nu}=2\int\frac{\alpha^2F(\omega)}{\omega}d\omega,
	\label{eq:lambda}
\end{equation}
where $\lambda_{\mathbf{q}\nu}$ is an electron-phonon coupling constant associated with a momentum $\mathbf{q}$ and  branch $\nu$. Similarly, a pocket-resolved electron-phonon coupling constant $\lambda(\text{P}_i\leftrightarrow\text{P}_j)$ can be obtained by replacing the $\alpha^2F(\omega)$ in Eq.~(\ref{eq:lambda}) with $\alpha^2F(\text{P}_{i}\leftrightarrow \text{P}_{j},\omega)$. The temperature dependent superconducting gaps and DOS are obtained with an anisotropic imaginary-time Migdal-Eliashberg formalism~\cite{Margine2013,Ponce2016} with subsequent analytic continuation to the real axis  using Pad$\acute{\textrm{e}}$ functions. The Matsubara frequency cutoff was set to 0.32 eV, which is nearly 9 times the maximum phonon  frequency in the system. Only the electronic states within the energy window of 0.1 eV around Fermi levels are involved and the convergence of the size of the energy window is confirmed as shown in Fig.~\ref{figs6:a2f_conv}(b) in the supplement. The Allen-Dynes-modified McMillan equation~\citep*{Allen1975,McMillan1968,Giustino2017} is also employed for the evaluation of $T_{c}$, where  an effective Coulomb potential $\mu_{\text{c}}^{*}$ is set to  be 0.15,  estimated according to the DOS at Fermi energy in the CDW phase~\cite{zheng2017charge}: $\mu_{\text{c}}^{*}\approx [0.26N_{\text{F}}/(1+N_{\text{F}})]$ \citep*{Garland1972,Xi2016}. 

\section{Convergence of the Eliashberg spectra}
\begin{figure}[H]
	\centering
	\includegraphics[width=140mm]{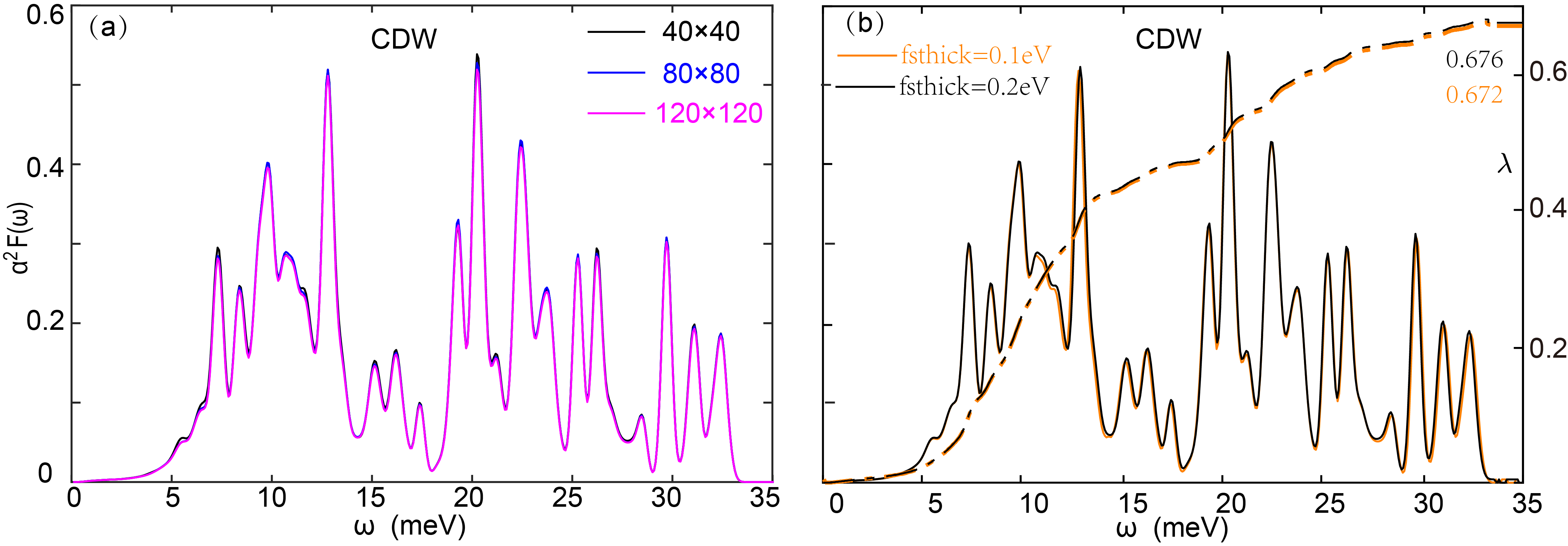} \\
	\caption{(a) Computed Eliashberg spectra of the CDW phase using both $\mathbf{k}$- and $\mathbf{q}$-grid of $40\times40$(black line), $80\times80$ (blue line) and $120\times120$ (pink line). (b) Computed Eliashberg spectra and electron-phonon coupling constants using ``fsthick'' = 0.1 and 0.2 eV, respectively.  The $\mathbf{k}$- and $\mathbf{q}$-grid used here are both $40\times40$. The broadening of electrons and phonons used to compute the spectra in (a) and (b) are 6.5 and 0.325 meV, respectively.
	}
	\label{figs6:a2f_conv}
\end{figure}

In the numerical simulations of Eliashberg spectra, each $\delta$ function related to electrons and phonons  [see Eq. (\ref{eq:a2f}) in the supplement] is replaced by a Gaussian function with an appropriate broadening. There are three $\delta$ functions in the Eq. (\ref{eq:a2f}), where $\delta(\epsilon_{\mathbf{k}}-\epsilon_{F})$ and $\delta(\epsilon_{\mathbf{k'}}-\epsilon_{F})$ are related to electrons, and $\delta(\omega-\omega_{\mathbf{k} - \mathbf{k'},\nu})$ is assigned to phonons. In this paper, the computed Eliashberg spectra are converged to small broadening of electrons (6.5 meV) and phonons (0.325 meV) with respect to the Brillouin zone samplings. As shown in Fig.~\ref{figs6:a2f_conv}(a) where the broadening of 6.5 (electrons) and 0.325 (phonons) meV are used to obtained Eliashberg spectra, when the grids denser, the resultant spectra are nearly identical, indicating the excellent convergence of broadening with respect to Brillouin zone samplings.  The convergence of the $\alpha^{2}F(\omega)$ with respect to the energy windows around Fermi energy is also confirmed (the parameter called ``fsthick'', is the size of a energy window within which the electronic states are taken into account). Fig.~\ref{figs6:a2f_conv}(b) displays the computed spectra using ``fsthick'' = 0.1 and 0.2 eV, respectively. The $\mathbf{k}$ and $\mathbf{q}$ grids are both $40\times40$. Again, the resultant spectra and electron-phonon coupling constants are nearly identical when ``fsthick'' is double.

\section{Influence of the CDW on  Eliashberg spectrum}

\begin{figure}[H]
	\centering
	\includegraphics[width=140mm]{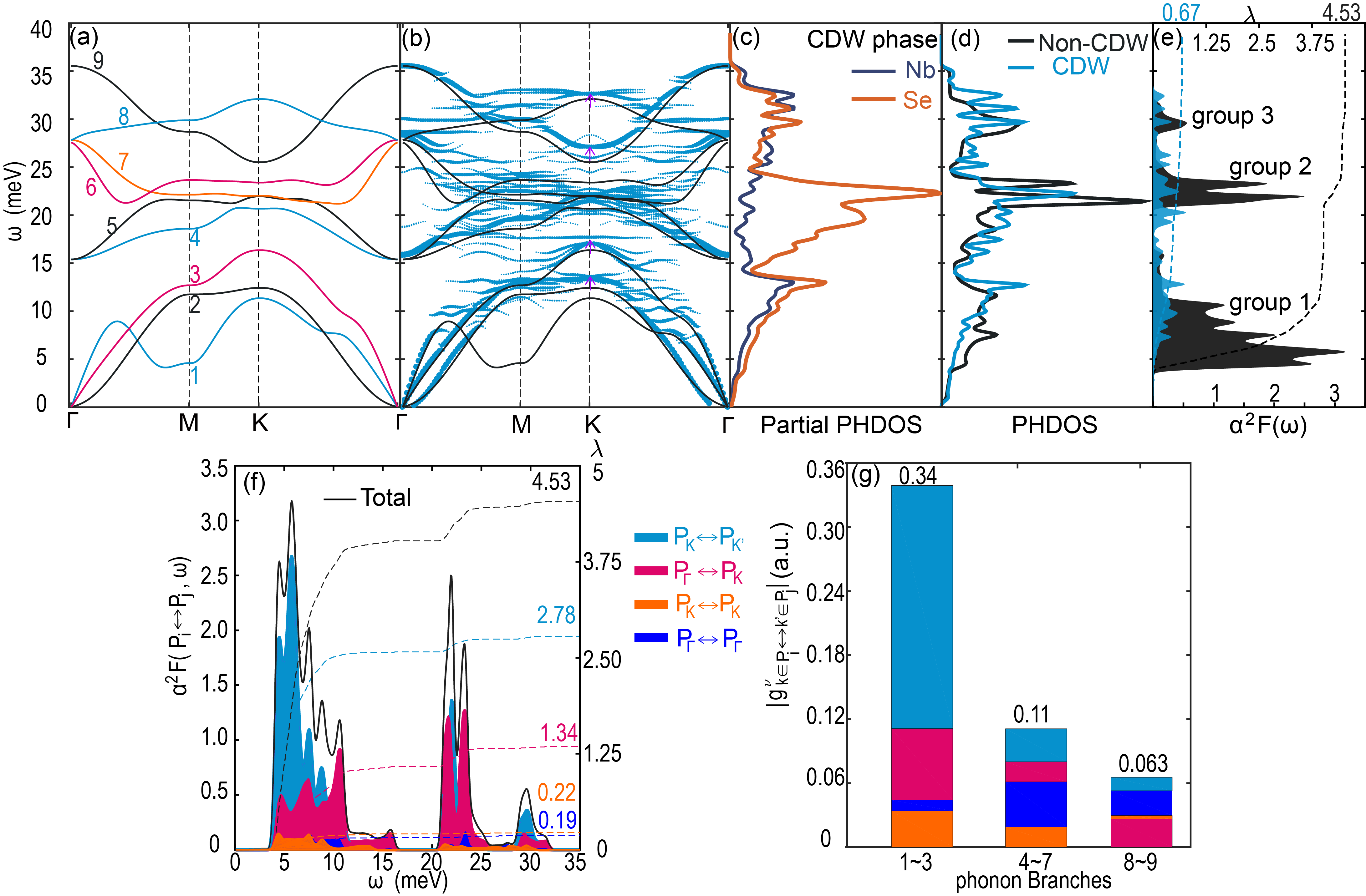} \\
	\caption{ (a) Phonon dispersions of the non-CDW phase with electronic broadening $\sigma = 0.03$  Ry. The phonon branches are marked by the Arabic numerals and colors. (b) unfolded phonon dispersion of the CDW phase (light-blue dots, $\sigma = 0.01$ Ry) in primitive Brillouin zone, combined with the dispersion of the non-CDW (black curves, $\sigma = 0.03$ Ry). The sizes of the  dots are positive correlated to the unfolding weights. The pink arrows label the blue shifts of the phonons. (c) Partial phonon density of states (PHDOS) of the CDW phase. (d) PHDOS of the non-CDW and CDW phases. (e) Eliashberg spectra $\alpha^2F(\omega)$ with integrated  electron-phonon coupling  constants $\lambda(\omega)$ (colored dashed lines) of the non-CDW and CDW phases. (f) Pocket-resolved Eliashberg spectra of the non-CDW phase, $\alpha^2F(\text{P}_{i}\leftrightarrow\text{P}_{j},\omega)$ ($i, j = \mathbf{\Gamma}, \mathbf{K}, \mathbf{K'}$) and the corresponding integrated electron-phonon coupling constants $\lambda(\text{P}_{i}\leftrightarrow\text{P}_{j},\omega)$.  (g) Distributions  of  the value of $|g_{\mathbf{k}\in\text{P}_{i},\mathbf{k'}\in\text{P}_{j}}^{\nu} |$ (arbitrary units).}
	\label{figS2:pho_a2f}
\end{figure}

Influence of the CDW on  electron-phonon coupling is now analyzed. Fig.~\ref{figS2:pho_a2f}(e) displays the Eliashberg spectra $\alpha^2F(\omega)$ of the non-CDW (black region) and CDW phase (light-blue region). The spectrum of the  non-CDW phase is separated into 3 groups according to their energies, located within the ranges 3.5-16.7, 20.4-26 and 28-32.2 meV, contributed by the phonons with branches 1-3, 4-7 and 8-9, respectively. In contrast, the corresponding phonon density of state does not exhibit this kind of separations [Fig.~\ref{figS2:pho_a2f}(d)]. This observation reflects the fact that the phonons of branches 4 and 5 [see Fig.~\ref{figS2:pho_a2f}(a)] couple weakly with electrons as shown in Fig.~4(a) in the main text, leading to a gap in  $\alpha^2F(\omega)$ between the 1st  and 2nd group of phonons, though their DOS is not necessarily small. Similarly, the separation between the 2nd  and  3rd groups of phonons in Fig.~\ref{figS2:pho_a2f}(e) is due to the weak coupling of the 9th branch phonon with electrons. The same reasoning explains also the observation that the phonon frequencies of the non-CDW phase extend up to nearly 36 meV [Fig.~\ref{figS2:pho_a2f}(b)], whereas the corresponding $\alpha^2F(\omega)$ vanishes for $\hbar\omega > 32.2$ meV [Fig.~\ref{figS2:pho_a2f}(e)].  

The most remarkable change in $\alpha^2F(\omega)$ is an  overall sharp decrease of its intensity when the CDW is formed, leading to an sharp reduction of $\lambda$ from 4.53 down to 0.67. Such a reduction can be understood as follows. As shown in  Fig.~\ref{figS2:pho_a2f}(f), the pocket-resolved Eliashberg spectra (their definitions can ben found in section S2) $\alpha^2F(\text{P}_{\mathbf{K}}\leftrightarrow\text{P}_{\mathbf{K'}},\omega)$ and $\alpha^2F(\text{P}_{\mathbf{\Gamma}}\leftrightarrow\text{P}_{\mathbf{K}},\omega)$  dominate the total spectrum, especially in the 1st group of phonon. The electronic states associated with $\alpha^2F(\text{P}_{\mathbf{K}}\leftrightarrow\text{P}_{\mathbf{K'}},\omega)$ and $\alpha^2F(\text{P}_{\mathbf{\Gamma}}\leftrightarrow\text{P}_{\mathbf{K}},\omega)$ on $\mathbf{K}$  and $\mathbf{K'}$ pockets  are mainly  located on the fully band-gapped sectors as shown in  Figs.~3(b) in the main text and Fig.~\ref{fig4:kk_connec}(c) in the supplement, respectively. Therefore,  $\alpha^2F(\omega)$  undergoes a sharp decrease due to the band-gap formation on the Fermi surface when CDW is formed. 

\section{Temperature-dependent superconducting gaps and density of states}

\begin{figure}[H]
	\centering
	\includegraphics[width=140mm]{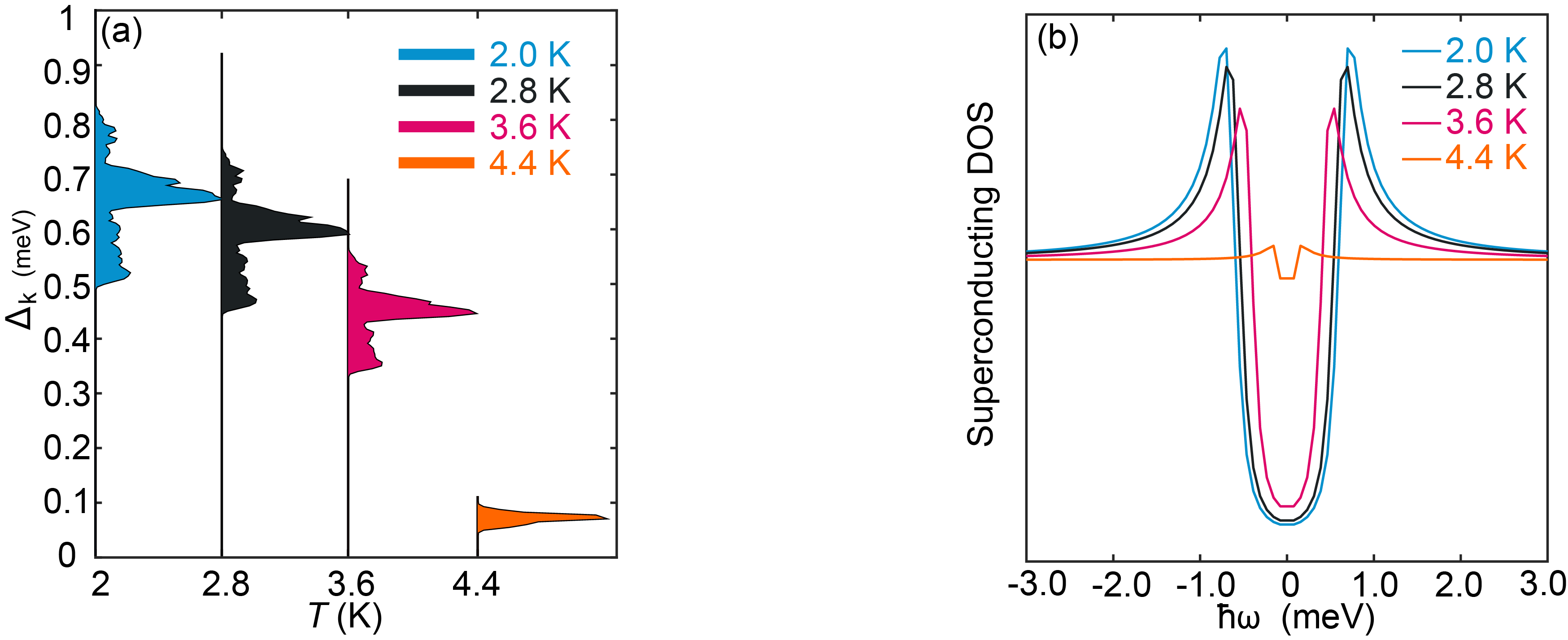} \\
	\caption{ (a) Computed anisotropic superconducting gap $\Delta_{\mathbf{k}}$ of the CDW phase around Fermi surface, evaluated as a function of temperature. For each temperature, the histograms are related to  the number of states on the Fermi surface with that superconducting gap energy. This plot is the same as Fig.~2(a) in the main text. (b) Computed superconducting density of states as a function temperature.}
	\label{figs9:sc_gap_dos}
\end{figure}

The computed superconducting density of states  also exhibit consistent trend with the computed superconducting gaps as temperature changes. As shown in Fig.~\ref{figs9:sc_gap_dos}(b), the superconducting density of state at 2 $K$ exhibit a peak at about 0.65 meV (blue curve). As temperature increases, the peaks of the curves suffer from gradual red shifts, attributable to the decrease of the superconducting gaps as shown in Fig~\ref{figs9:sc_gap_dos}(a).

\section{Two-gap feature in non-CDW monolayer NbSe$_{2}$}

\begin{figure}[H]
	\centering
	\includegraphics[width=140mm]{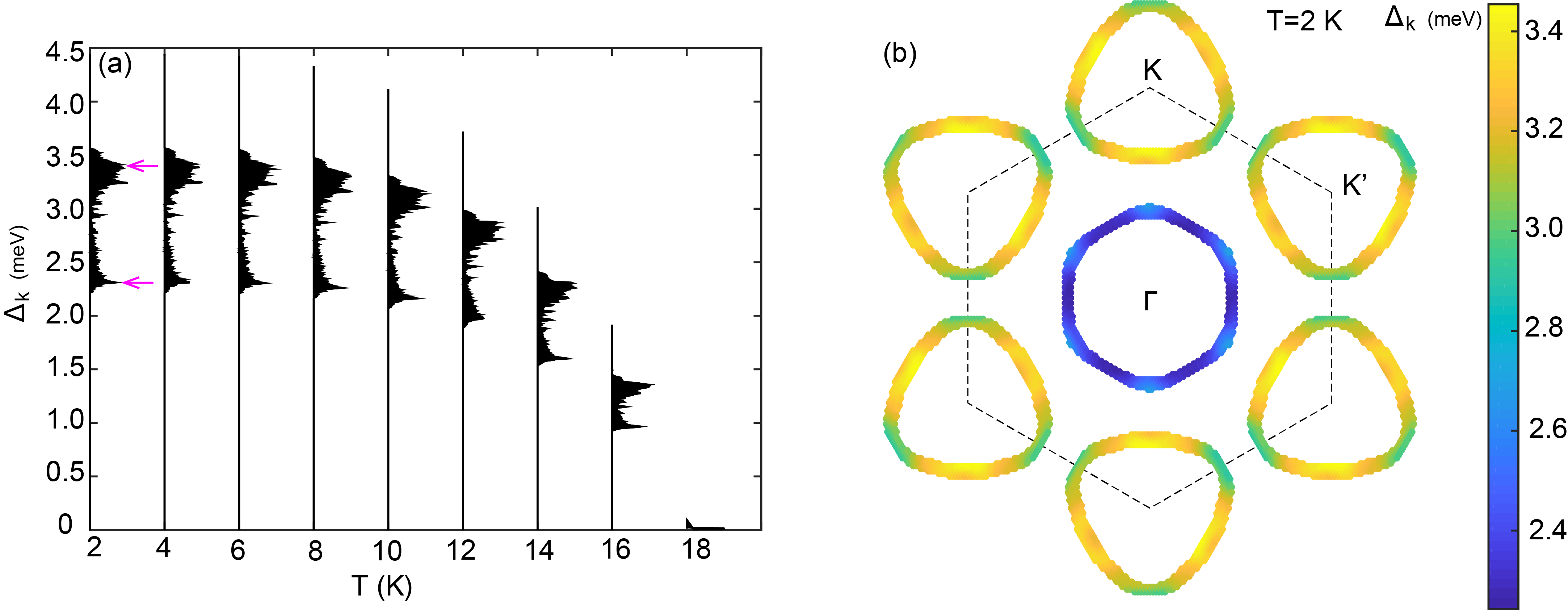} \\
	\caption{(a) Computed anisotropic superconducting gap $\Delta_{\mathbf{k}}$ of the non-CDW phase around Fermi surface, evaluated as a function of temperature. (b) Distribution of $\Delta_{\mathbf{k}}(T=2 K)$ of the non-CDW phase in primitive Brillouin zone.}
	\label{figs10:twogap}
\end{figure}

Fig.~\ref{figs10:twogap}(a) shows the computed superconducting gaps $\Delta_{\mathbf{k}}$ of the non-CDW phase around Fermi surface as a function of temperature. The two-gap feature is clearly seen. For example, when $T = 2~K$, a large gap centered at about 3.3 meV and a small one  at about 2.3 meV can be found. The positions of the above gaps are labeled by two pink arrows as shown in Fig.~\ref{figs10:twogap}(a). It should be mentioned that the computed superconducting transition temperature $T_{c} \approx 18~K$ and $\Delta_{\mathbf{k}}$ are much larger than experiment due to the neglect of CDW, leading to three intact Fermi pockets (not gapped by CDW) at $\mathbf{\Gamma}, \mathbf{K}$ and $\mathbf{K'}$. Further visualization of the distribution of $\Delta_{\mathbf{k}}$ for the non-CDW phase at $T = 2~K$ is shown in Fig.~\ref{figs10:twogap}(b). It is found that the relatively large superconducting gaps are contributed by the $\mathbf{K}$ and $\mathbf{K'}$ pockets, while the small ones are by the $\mathbf{\Gamma}$ pocket. The electronic states contributed to the $\mathbf{K}$ and $\mathbf{K'}$ pockets are found to be dominated by Nb-derived in-plane orbitals of 4$d_{x^2-y^2}$ + 4$d_{xy}$, and the electronic states related to the $\mathbf{\Gamma}$ pocket are mainly from the Nb-derived out-of-plane orbitals of 4$d_{z^2}$, which is discussed in section S7 of the supplement.

\section{Electronic characters of the three Fermi pockets}
\begin{figure}[H]
	\centering
	\includegraphics[width=140mm]{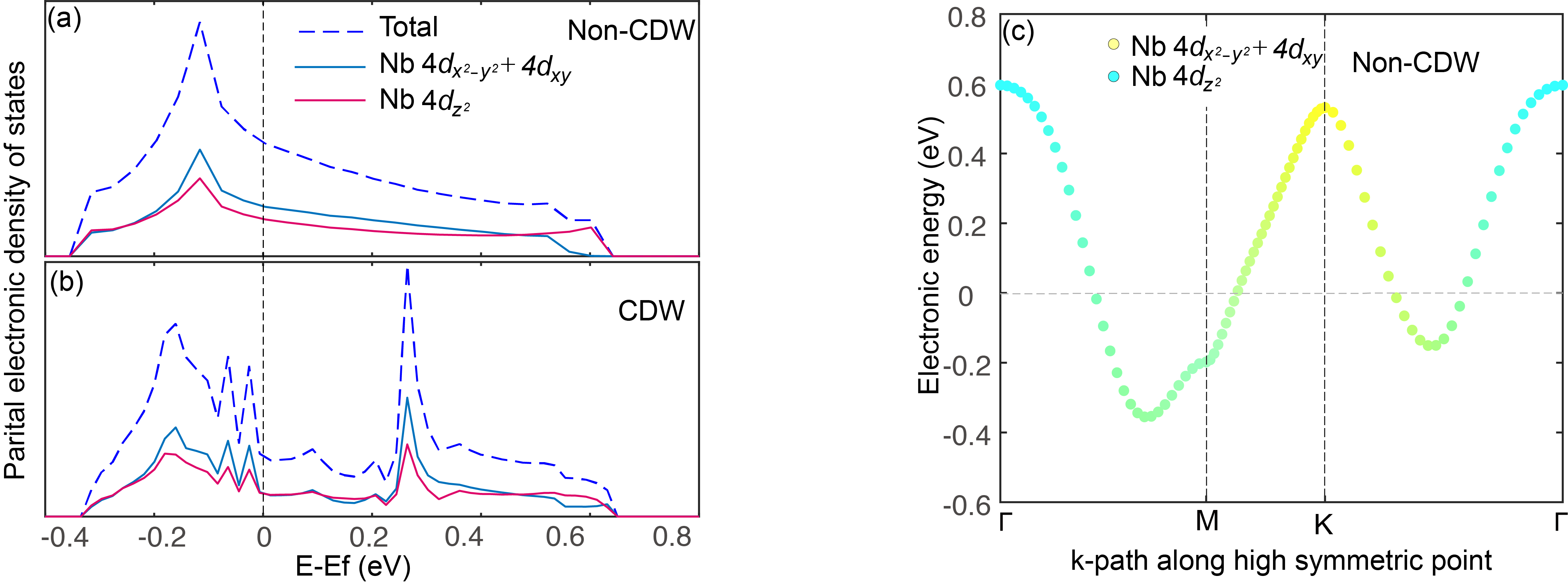} \\
	\caption{Total and projected density of states of the non-CDW (a) and CDW phases (b) around Fermi energy. (c) The projected bandstructure of the non-CDW phase onto $4d_{z^{2}}$, $4d_{x^2-y^2}$ and $4d_{xy}$-like orbitals of Nb atoms, respectively.}
	\label{figs7:projected-elec}
\end{figure}

The total and projected density of states [Figs.~\ref{figs7:projected-elec}(a) and \ref{figs7:projected-elec}(b)], and the projected bandstrcuture [Fig.~\ref{figs7:projected-elec}(c)] along high symmetric points of primitive Brillouin zone together show that, the electronic states at Fermi pocket $\mathbf{\Gamma}$ are dominated by the out-of-plane  orbitals of Nb-derived $4d_{z^2}$, while the ones at $\mathbf{K}$ and $\mathbf{K'}$ pockets are by the mixing of in-plane orbitals of Nb-derived $4d_{x^2-y^2}$ and $4d_{xy}$.

\section{The Brillouin zone unfolding scheme}
Eigenstates in the Brillouin zone of the CDW supercell, i.e. $|\mathbf{K}J>$, $|\mathbf{Q}\nu>$ (electronic wavefuction, phonon polarization vectors, respectively)  are unfolded to the Brillouin zone of the primitive cell of the non-CDW phase based on the method of unfolding technique~\citep*{Allen2013}. In light of this method, an eigenstate (take $|\mathbf{K}J>$ for example)  in supercell Brillouin zone  can be unfolded to primitive Brillouin zone with a weight function $W_{\mathbf{K}J}(\mathbf{G}) = 1/N\sum_{j = 1}^{N}\langle \mathbf{K} J | \hat{T}(\mathbf{r}_j) | \mathbf{K} J \rangle e^{-i(\mathbf{K} + \mathbf{G})\cdot \mathbf{r}_j}$ which measures the degree of the Bloch symmetry belonging to a primitive cell, that the eignstate of a supercell posesses.  The supercell contains $N$ primitive cells.  $| \mathbf{K} J \rangle $ donates an eigenstate of the supercell with  a band index $J$ and a wavevector $\mathbf{K}$.  $\mathbf{G}$ is a reciprocal lattice vector of the supercell, connecting $\mathbf{K}$ to a wavevector $\mathbf{k}$  in the primitive Brillouin zone in terms of $\mathbf{k} = \mathbf{K} + \mathbf{G}$. $\hat{T}(\mathbf{r}_j)$ is the translation corresponding to the $j$th primitive cell in the supercell.

\section{Pronounced $\lambda_{\mathbf{q}}$ around $\mathbf{q_{\text{CDW}}}$} 
\begin{figure}[H]
	\centering
	\includegraphics[width=140mm]{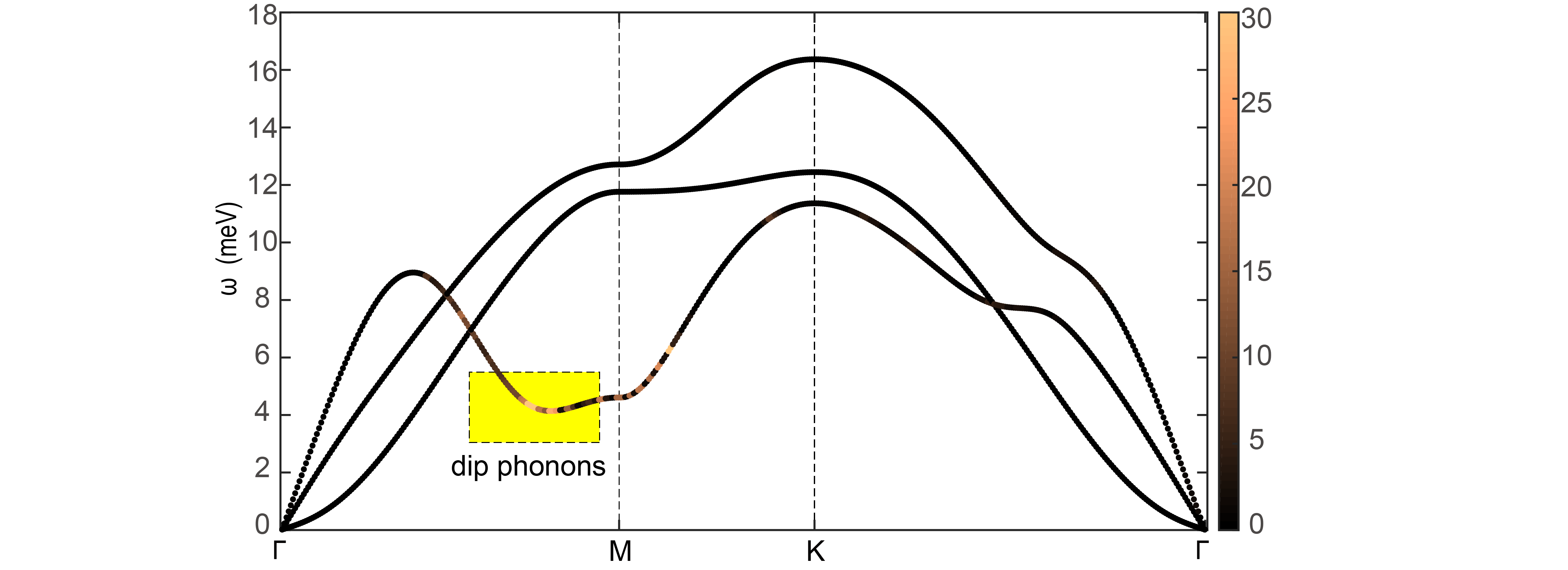} \\
	\caption{ Phonon dispersion of the three acoustic branches, $\omega_{\mathbf{q}\nu=1-3}$, of the non-CDW phase. The color code depicts the dimensionless electron-phonon coupling constant $\lambda_{\mathbf{q}\nu}$. This figure is similar with Fig.~4(a) in the main text, but with a different colorbar, showing the pronounced electron-phonon coupling strength of the dip phonons around $\mathbf{q_{\text{CDW}}}$. 
	}
	\label{figs4:eph}
\end{figure}

\section{qualitative analysis  of electron-phonon coupling of triangle* CDW phase }

\begin{figure}[H]
	\centering
	\includegraphics[width=140mm]{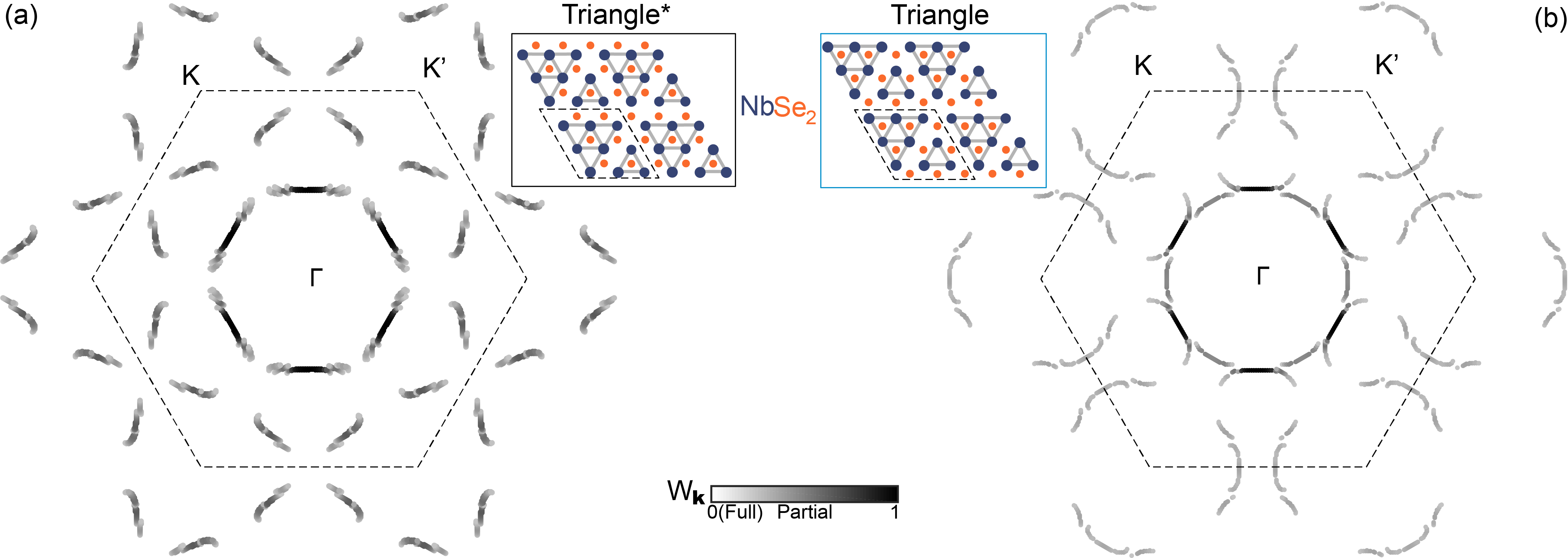} \\
	\caption{ Unfolded Fermi surfaces of the triangle* CDW phase (a) and triangle CDW phase (b). The latter is the same as Fig.~2(b) in the main text .  The insets display the corresponding crystal structures from topview.}
	\label{figs6:unfoldFS}
\end{figure}

The electron-phonon coupling of the triangle* CDW phase is qualitatively  analyzed by computing unfolded Fermi surface, since electron-phonon coupling is closely related to the electronic states at Fermi energy. Fig.~\ref{figs6:unfoldFS}(a) displays the unfolded Fermi surface of the triangle* CDW phase and its crystal structure from topview. The same information of the triangle CDW phase is also shown in Fig.~\ref{figs6:unfoldFS}(b) for comparison. It can be seen that the Fermi surface of the triangle CDW phase suffer from more extensive gapped after CDW, since the unfolding weights (W$_{\mathbf{k}}$) are overall smaller than the case of the triangle* CDW phase [comparing Fig.~\ref{figs6:unfoldFS}(b) with ~\ref{figs6:unfoldFS}(a)]. Therefore, it is expected that the electron-phonon coupling in the triangle* CDW phase should be stronger than that of the triangle CDW phase. This is indeed the case as reported by~\citep{Lian2018}, where the electron-phonon coupling constant of the ``3+6-filled'' CDW phase (triangle* phase here) is larger than that of the ``3+6-hollow'' CDW phase (triangle phase here).  The same reasoning explains also the computed superconducting transition temperature of the triangle* CDW [7.6 $K$ in the work~\citep{Lian2018}] is larger than that of triangle phase [4.5 $K$ in the work~\citep{Lian2018}].


\section{Visualization of the selected electron-phonon coupling matrix elements }

\begin{figure}[H]
	\centering
	\includegraphics[width=160mm]{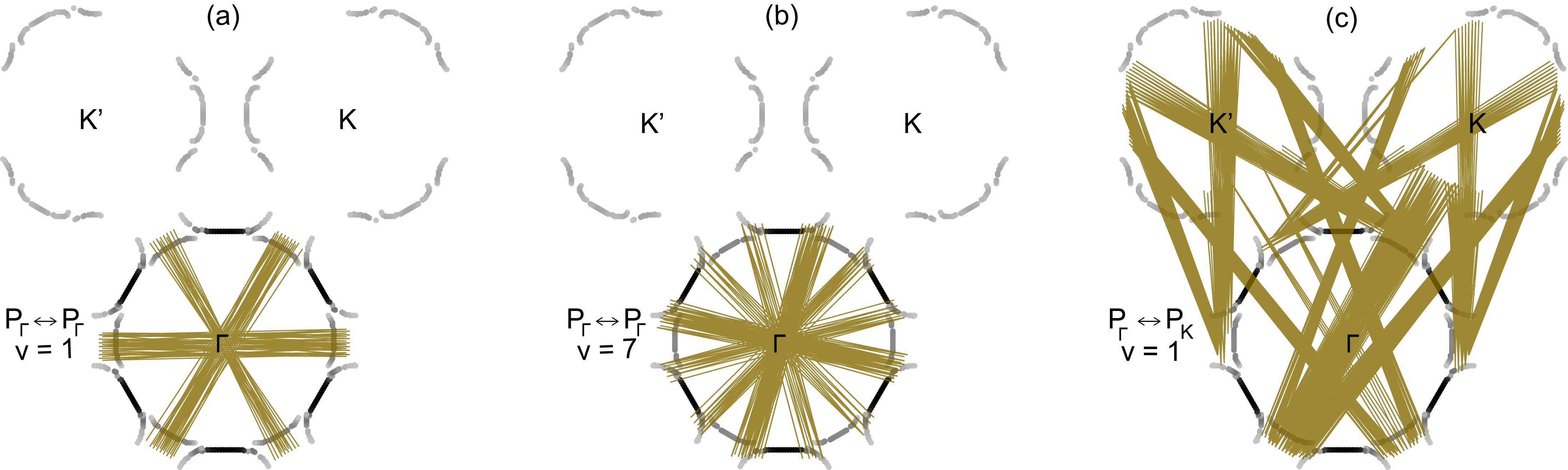} \\
	\caption{ Visualization of  pocket-resolved matrix elements $\left |g_{\mathbf{k} \in \text{P}_{\mathbf{\Gamma}},\mathbf{k'} \in\text{P}_{\mathbf{\Gamma}}}^{\nu=1} \right | $ (a), $\left |g_{\mathbf{k} \in \text{P}_{\mathbf{\Gamma}},\mathbf{k'} \in\text{P}_{\mathbf{\Gamma}}}^{\nu=7} \right | $ (b), $\left |g_{\mathbf{k} \in \text{P}_{\mathbf{\Gamma}},\mathbf{k'} \in\text{P}_{\mathbf{K}}}^{\nu=1} \right | $ (c). Those elements with dominant values for the above four types matrix elements are first picked out. Each pair of $\mathbf{k}$, $\mathbf{k'}$ related to the selected matrix elements is subsequently located and visualized by drawing the connecting vector $\mathbf{q} = \mathbf{k'} - \mathbf{k}$ using green line as shown in (a-c).
	}
	\label{fig4:kk_connec}
\end{figure}


\end{document}